\documentclass{aa}  

\usepackage{graphicx}

\usepackage{txfonts}
\usepackage{units}
\usepackage{soul}
\setstcolor{red}
\usepackage{keyval}

\usepackage{natbib}
\usepackage{hyperref} 
\usepackage{xcolor}
\usepackage{xspace}

\newcommand{\emos}{EMOs\xspace}
\newcommand{\ttar}{Sh2-106\xspace}

\newcommand{\as}[2]{$#1\mbox{$''\mskip-7.6mu.\,$}#2$}

\newcommand{\ujpb}{{$\mu$Jy\,beam$^{-1}$}\xspace}

\newcommand{\hii}{H\,\textsc{ii}\xspace}

\defcitealias{Dzib2015}{D15}

\begin{document}

\title{
Distance to Sh2-106 from Gaia DR3 and its embedded radio population: implications for a candidate explosive outflow
}

\titlerunning{Gaia distance and radio sources in Sh2-106}

\author{
  Sergio A. Dzib\inst{1} 
}

\institute{
  Max-Planck-Institut f\"ur Radioastronomie, Auf dem H\"ugel 69, D-53121 Bonn, Germany\\
  \email{sdzib@mpifr-bonn.mpg.de}
}

\date{Received ; accepted }

\abstract{
Sh2-106 has recently been proposed as a candidate explosive molecular outflow (EMO), but the physical interpretation of the region depends critically on its distance. Published estimates span a wide range, leading to large uncertainties in the inferred size, energetics, and evolutionary timescale of the system. Using {\it Gaia} DR3 astrometry, we identify a kinematically coherent stellar population associated with Sh2-106 and derive a cluster parallax of $\varpi_{\rm corr}=0.607\pm0.013$\,mas, corresponding to a distance of $1.65\pm0.04$\,kpc. This value is significantly larger than the commonly adopted extinction-break estimate of 1.09\,kpc. At this revised distance, the inferred kinetic energy of the expanding ionized nebula increases by a factor of $\sim6.5$, reaching $E_{\rm exp}\simeq1.3\times10^{48}$\,erg and placing Sh2-106 in the same order-of-magnitude energetic regime as the Orion BN/KL explosive event, although at a substantially older dynamical age ($\sim3500$\,yr). Archived 5.8\,GHz Karl G. Jansky Very Large Array observations reveal ten compact radio sources in the central region, identifying embedded stellar objects that are suitable for future multi-epoch radio astrometry. No unambiguous high-velocity stellar ejecta are detected in {\it Gaia} DR3, although S106\,IR shows a modest peculiar transverse velocity of $\sim5$\,km\,s$^{-1}$ relative to the cluster centroid. The Gaia-based cluster distance, therefore, significantly revises the physical scale and energetics of Sh2-106 and provides the observational framework required to test whether the region represents an older analogue of the Orion BN/KL dynamical disintegration or a distinct explosive phenomenon.
}

\maketitle

\section{Introduction}

Explosive molecular outflows (\emos) represent one of the most dramatic and least understood phenomena associated with massive star formation. In contrast to the collimated bipolar outflows driven by disk accretion \citep{zapata2017}, these events appear as nearly isotropic networks of high-velocity, finger-like filaments emanating from a common origin \citep{bally2011}. Their kinetic energies can exceed those of standard protostellar outflows by orders of magnitude, suggesting a fundamentally different physical origin.

The Orion BN/KL region ($d\sim0.39$\,kpc; \citealt{dzib2026}) provides the clearest example of such an event \citep{bally2011}. There, dozens of molecular filaments trace a roughly radial expansion pattern, while the proper motions of several stars converge to a common origin \citep[e.g.,][]{gomez2008,luhman2017,rodriguez2020}. These observations support a scenario in which the dynamical decay of a compact multiple system expelled both stars and circumstellar material \citep{bally2011}. Whether Orion represents a unique case or the nearest example of a broader class of explosive events remains an open question.

Several additional \emos candidates have been proposed, including DR\,21 \citep{zapata2013}, G5.89--0.39 \citep{zapata2020}, Sh2--106 \citep{bally2022}, IRAS~16076--5134 \citep{guzman2022}, IRAS~12326--6245 \citep{zapata2023}, and G34.26+0.15 \citep{issac2025}. In any of these cases, however, the stellar dynamical component of the explosion has been directly tested, largely because these regions lie at distances greater than 1\,kpc and are heavily obscured.

The bipolar \hii\ region Sh2-106 has recently emerged as one of the most promising candidates. High-resolution molecular observations reveal a system of high-velocity filamentary streamers with a quasi-radial distribution and a Hubble-like expansion pattern comparable to that observed in Orion BN/KL \citep{bally2022}. Previous analyses suggest that the energetics of the expanding nebula may approach the $\sim10^{48}$\,erg regime, although these estimates depend sensitively on the adopted distance to the region.

Published distance estimates to Sh2-106 span a significant range, from $1.09\pm0.05$\,kpc \citep{zucker2020} to $1.3\pm0.1$\,kpc \citep{xu2013}. The massive young stellar object S106\,IR, located at the center of the nebula, has a {\it Gaia} DR3 parallax of $\varpi = 0.56 \pm 0.13$\,mas \citep{gaia2023}, corresponding nominally to $1.8\pm0.4$\,kpc. However, the large astrometric excess noise (1.1\,mas) indicates that the single-object measurement may be unreliable. Although Sh2-106 has long been known as a young embedded star-forming region, no Gaia-based astrometric population has yet been explicitly linked to S106\,IR. Because the physical size, energy, and dynamical timescales of the outflow scale directly with distance, a robust determination of the distance to Sh2-106 is essential for assessing the nature of the event.

In this paper, we revisit the distance to Sh2-106 using {\it Gaia} DR3 astrometry of the surrounding stellar population. By identifying a kinematically coherent group of stars in the region, we derive a cluster-based distance that is less sensitive to the astrometric uncertainties of individual objects. In addition, we present new 5.8\,GHz Karl G. Jansky Very Large Array (VLA) observations at sub-arcsecond resolution to identify compact radio sources associated with deeply embedded young stellar objects. These observations provide a complementary view of the stellar content of the region and help establish an observational framework for evaluating the proposed explosive outflow scenario.

\begin{figure*}
    \centering
    \includegraphics[width=0.99\linewidth]{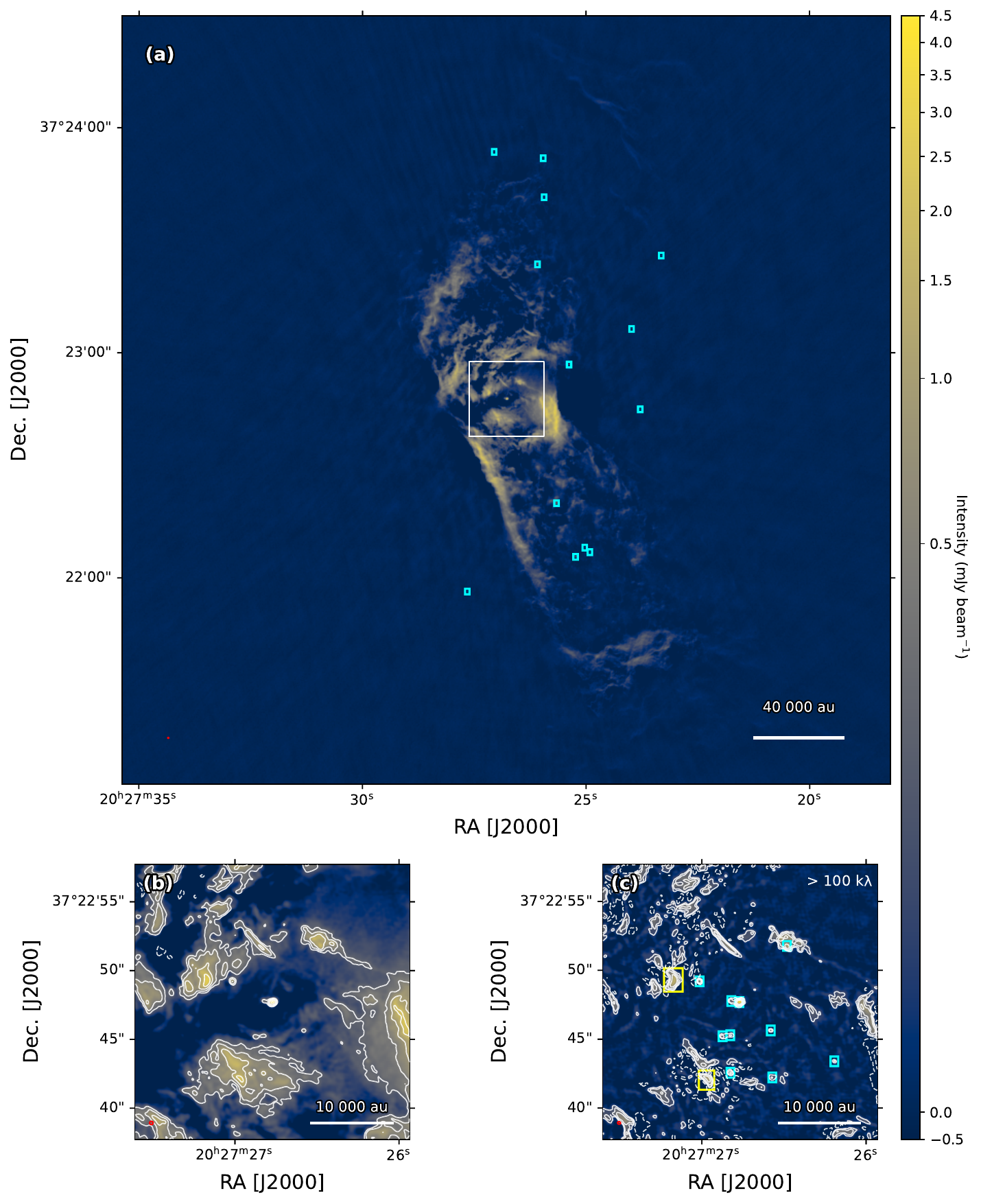}
    \caption{VLA 5.7\,GHz continuum image of the \object{Sh2-106} region, centered on the massive star S106~IR. 
(a) Full-field image. 
(b) Zoom-in on the core region (white square in panel (a)) showing both compact and extended emission. 
(c) Same field as in panel (b), imaged using only visibilities with $uv$-distances $>100$\,k$\lambda$, highlighting the compact structures. The cyan squares mark positions of compact sources in the core (panel (c)) and outside the core (panel(a)). The yellow squares in panel (c) indicate two additional radio sources with bow-shock morphology. 
All panels share the same color scale, shown at right, in units of mJy\,beam$^{-1}$. 
Contours in panels (b) and (c) are drawn at levels of $-5$, 5, 10, 20, 40, and 80 times the rms noise in the central region (48 and 19\,$\mu$Jy\,beam$^{-1}$ for the images without and with the $uv$-cut, respectively). 
The synthesized beam is shown as a red ellipse in the lower-left corner of each panel. 
Linear scale bars are shown assuming a distance of 1.65\,kpc (see Sects. \ref{sect:gaiad} and \ref{sec:dist}).}
    \label{fig:1}
\end{figure*}

\section{Observations and Data}\label{sec:data}

\subsection{Gaia}

We retrieved astrometric data from the ESA Gaia Archive\footnote{https://gea.esac.esa.int/archive/} using the Gaia Data Release 3 (DR3) whose content is described in \citet{gaia2023}. Sources were selected within a circular region of radius $5'$ centered on the nominal position of S106\,IR. We found 362 sources within these constraints. 

We extracted the positions, parallaxes ($\varpi$), proper motions ($\mu_{\alpha}^*, \mu_\delta$), and their associated uncertainties. To ensure reliable astrometry, we restricted the sample to sources with renormalized unit weight error (RUWE) $< 1.4$. No prior selection was applied on parallax sign or distance.

\subsection{VLA observations}
The archived observations were obtained with the VLA of the National Radio Astronomy Observatory (NRAO) in its A configuration on July 30, 2023, under project code 23A-152. The data were acquired in C-band (4--8\,GHz) using the 3-bit sampler, which provides simultaneous wide-band continuum sensitivity and high spectral resolution. In this work, we focus on the semi-continuous continuum data; the high-spectral-resolution dataset will be presented in a future study.

For the continuum setup, the band was divided into 16 spectral windows (SPWs), each 128\,MHz wide and composed of 128 channels of 1\,MHz. Two frequency ranges were covered: 4.4--5.4\,GHz and 6.0--7.0\,GHz. The total observing time was three hours. The flux-density scale was set using J1331+3030 (3C\,286), and phase calibration was performed using interleaved observations of the nearby gain calibrator J2015+3710, with a cycle of 1 minute on the calibrator and 7 minutes on the target.

The data were calibrated and edited using the standard NRAO calibration pipeline within the Common Astronomy Software Applications package \citep[CASA;][]{casa2022}. The pipeline performs automated flagging of corrupted data and applies delay, bandpass, flux-density, and complex gain calibrations.\footnote{A detailed description of the VLA pipeline is available at \url{https://science.nrao.edu/facilities/vla/data-processing/pipeline}.}

Initial imaging of the calibrated visibilities revealed that the field is dominated by bright, spatially extended emission. As a consequence, the achieved root-mean-square (rms) noise level exceeded $100\,\mu$Jy\,beam$^{-1}$, in the central area where more of the extended emission is present, and $\sim20\,\mu$Jy\,beam$^{-1}$ in areas free of emission. These values are significantly above the expected theoretical sensitivity of $\sim3.5\,\mu$Jy\,beam$^{-1}$. This discrepancy indicated that residual phase errors were limiting the dynamic range.

To mitigate these effects, we performed phase-only self-calibration on the target field. An initial model was constructed by imaging the data with multi-scale deconvolution and defining clean masks around the brightest emission features. Using this model, phase solutions were derived and applied to the visibilities. The imaging and calibration steps were iterated, progressively refining the sky model and reducing residual phase errors. The solution intervals were chosen to preserve adequate signal-to-noise while tracking time-dependent atmospheric variations. 

After self-calibration, the image quality improved substantially (panel (a) in Fig.~\ref{fig:1}). The final rms noise level is $4.4\,\mu$Jy\,beam$^{-1}$ in emission-free outer regions and $\sim50\,\mu$Jy\,beam$^{-1}$ in the central area, where extended emission dominates. The final image after self-calibration was constructed with pixel sizes of \as{0}{08}, and an image size of 2560 pixels per side using a weighting scheme intermediate between natural and uniform (\texttt{robust=0}). The synthesized beam of the final image is \as{0}{27}\,$\times$\,\as{0}{25} at a position angle of 83$^\circ$. We verified that the brightest compact sources detected in the final image were also present in the pre-self-calibrated maps, ensuring that no artificial structures were introduced during self-calibration. 

To better study the compact radio sources, we produced a series of additional images by removing baselines with ($u,\,v$)$<$100\,k$\lambda$, effectively removing emission from structures with sizes $\gtrsim2''$. Three such additional images were generated: a full-band (FB) image including the full 2\,GHz bandwidth, a low-sideband (LSB; 4.4--5.4\,GHz) image, and an upper-sideband (USB; 6.0--7.0\,GHz) image. A Stokes~V image was also produced; however, it contains no significant emission and was not used for further analysis. All the additional images were constructed with a pixel size of \as{0}{05}, 1280 pixels per side, and Briggs weighting with \texttt{robust=$-1.0$}. Properties of final used images are summarized in Table~\ref{tab:obs}.

 \begin{table}
\small
\caption{VLA observations and image properties.}
    \begin{tabular}{ccccc}
         \hline
         \hline
 Sub- & $uv$-cut &{Synthesized beam}& $\sigma_{noise}$-central & $\sigma_{noise}$-outer  \\
band  & (k$\lambda$) &($\theta_{\rm maj.}\times\theta_{\rm min.}$; P.A.) & (\ujpb) & (\ujpb)\\
\hline
 FB & ...  &\as{0}{27}$\times$\as{0}{25}; $+83^{\circ}$  & 48 & 4.4  \\
 FB &$>100$&\as{0}{23}$\times$\as{0}{20}; $+76^{\circ}$  & 19 & 6.5  \\
LSB &$>100$&\as{0}{29}$\times$\as{0}{25}; $+76^{\circ}$  & 23 & 9.7  \\
USB &$>100$&\as{0}{23}$\times$\as{0}{20}; $+73^{\circ}$  & 17 & 9.2  \\
         \hline
    \end{tabular} \label{tab:obs}
\end{table}

\section{Analysis methods and results}\label{sec:results}

\subsection{Gaia astrometric analysis}\label{sect:gaiad}

Recent Gaia-based catalogs of stellar clusters and associations \citep[e.g.,][]{castro2022,hunt2023} do not report a cluster associated with Sh2-106. We therefore searched for a kinematically coherent stellar population in the region using Gaia DR3 data. Our sample consists of Gaia DR3 sources within a circular region of radius $5'$ centered on S106\,IR. We used the Gaia astrometric parameters $(\varpi,\mu_{\alpha}^*,\mu_\delta)$ to identify kinematically coherent stellar populations. 

Clustering was performed using the DBSCAN algorithm in the five-dimensional parameter space defined by sky position $(\alpha,\delta)$, parallax ($\varpi$), and proper motions $(\mu_{\alpha}^*,\mu_\delta)$. All parameters were standardized to zero mean and unit variance prior to clustering to ensure comparable scaling among dimensions. The DBSCAN parameters were set to $\epsilon = 0.6$ and \texttt{min\_samples}=5. We explored $\epsilon$ values between 0.4 and 0.8 and \texttt{min\_samples} between 4 and 8. The cluster centroid and mean parallax remain stable within this range, indicating that the cluster identification is insensitive to the choice of parameters.

\begin{figure}
    \centering
    \includegraphics[width=1.0\linewidth]{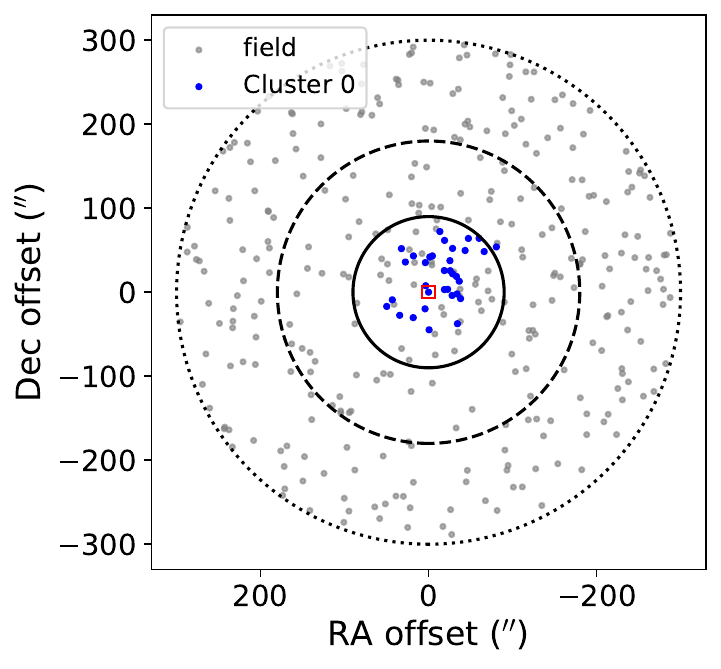}
    \caption{Sky distribution of Gaia sources within $5'$ of Sh2-106. Sources identified as cluster members by DBSCAN with a selection radius of $3'$ are shown in blue. Field stars are shown in gray. The members form a clear spatial overdensity centered near S106\,IR (inside the red square at the coordinate center).}
    \label{fig:g1}
\end{figure}

\begin{figure}
    \centering
    \includegraphics[width=1.0\linewidth]{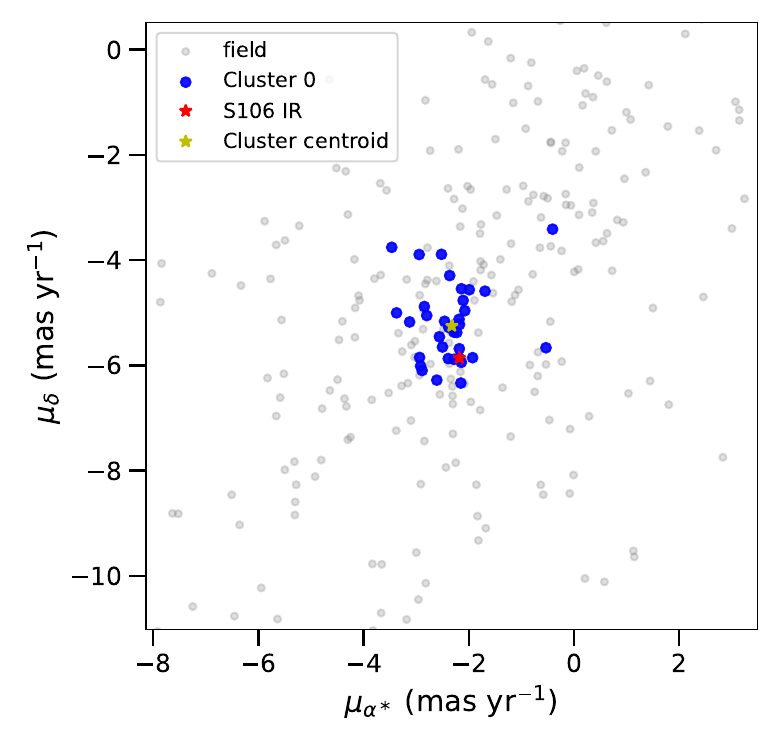}
    \caption{Proper motion distribution of Gaia sources within $5'$ of Sh2-106. Sources identified as cluster members by DBSCAN (when $r$ is set to $3'$) are shown in blue and S106\,IR as a red star, while field sources are shown in grey. The cluster members form a tight kinematic overdensity in proper-motion space.}
    \label{fig:g2}
\end{figure}

To evaluate the robustness of the clustering and the impact of field contamination, the analysis was repeated for circular regions of radius $1.5'$, $3.0'$, and $5.0'$ centered on S106\,IR, containing 68, 144, and 362 sources, respectively. In all cases, the algorithm consistently identifies a dominant overdensity (Cluster~0) with a stable proper-motion centroid (Table~\ref{tab:gaia}). The spatial distribution and proper-motion properties of the cluster members are shown in Figs.~\ref{fig:g1} and~\ref{fig:g2}.

The resulting astrometric properties of Cluster 0 are summarized in Table~\ref{tab:gaia}. The parallaxes derived within $1.5'$ and $3'$ are consistent within uncertainties, but the smaller sample contains only 11 sources and therefore provides a less stable estimate of the cluster parallax. The $3'$ sample increases the number of members to 34 while maintaining a similar mean parallax and proper-motion centroid, indicating that contamination remains limited at this radius. At larger radii ($5'$), the systematic increase in both the mean parallax and its dispersion suggests significant contamination by foreground stars (see Appendix A). We therefore adopt the $3'$ sample as the best compromise between statistical robustness and minimal field contamination.

\begin{figure}
    \centering
    \includegraphics[width=0.8\linewidth]{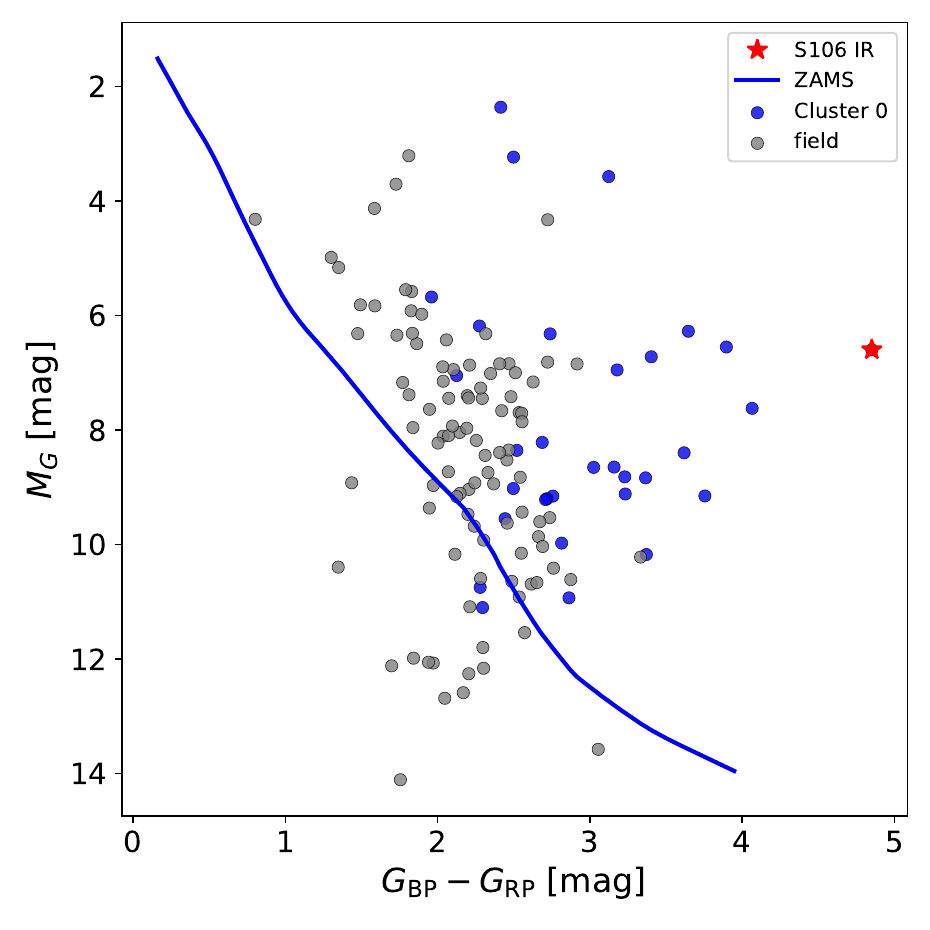}
    \caption{Gaia color–magnitude diagram of sources in the vicinity of Sh2-106. The blue curve shows the zero-age main sequence in the Gaia photometric system. }
    \label{fig:gcmd}
\end{figure}

Figure~\ref{fig:gcmd} shows the Gaia color--magnitude diagram of the selected cluster members and the surrounding field population. Most candidate members are located above the zero-age main sequence and extend toward redder colors, as expected for young stars seen through substantial extinction. Additional support for the youth of the population comes from a SIMBAD cross-match: nine members are classified as YSOs, five as YSO candidates, and one as an X-ray source. The remaining 19 objects are listed simply as stars, without a more specific classification. Given the strong and spatially variable extinction toward Sh2-106, we restrict ourselves to a qualitative interpretation of the diagram and do not attempt a formal isochronal age analysis.

Using the weighted mean parallax within $3'$, $\varpi = 0.590 \pm 0.008$\,mas, and applying a Gaia DR3 zero-point correction of $+0.017\pm0.010$\,mas \citep{lindegren2021} yields $\varpi_{\rm corr} = 0.607\pm0.013$\,mas. The quoted uncertainty includes the statistical error of the weighted mean and an estimate of the Gaia systematic zero-point uncertainty added in quadrature. This corresponds to a distance of $1.65 \pm 0.04$\,kpc.

\begin{table}
\setlength{\tabcolsep}{2.5pt}
\caption{Astrometric properties of Cluster 0 derived from \texttt{DBSCAN} clustering using different selection radii.}
\label{tab:gaia}
\centering
\begin{tabular}{cccccc}
\hline\hline
Radius & $N$ & $\langle \varpi \rangle$ & $\sigma_{\varpi}$ & $\langle \mu_{\alpha*} \rangle$ & $\langle \mu_{\delta} \rangle$ \\
($'$) &  & (mas) & (mas) & (mas yr$^{-1}$) & (mas yr$^{-1}$) \\
\hline
1.5 & 11 & $0.569 \pm 0.018$ & 0.166 & $-2.37$ & $-5.36$ \\
3.0 & 34 & $0.590 \pm 0.008$ & 0.291 & $-2.35$ & $-5.20$ \\
5.0 & 71 & $0.692 \pm 0.006$ & 0.354 & $-2.34$ & $-5.00$ \\
\hline
\end{tabular}
\end{table}

\subsection{Radio images and compact radio sources}

The full-band image without a $uv$-cut recovers the extended ionized emission associated with the bipolar \hii\ region (Fig.~\ref{fig:1}a). The large-scale morphology clearly traces the well-known bipolar structure of Sh2-106. The image resolves the filamentary structure of the ionized gas in greater detail than previous VLA observations \citep[e.g.,][]{bally2022}, revealing morphology closely resembling that seen in the HST/WFC3 F658N ([N\,{\sc ii}]) image presented by \citet{bally2022}.

To isolate compact emission, we produced images excluding baselines with $(u,\,v) < 100$\,k$\lambda$. This filtering suppresses structures larger than $\sim2''$ and enhances compact sources (Fig.~\ref{fig:1}c). By visually inspecting the map, we identified 23 compact radio sources within the imaged area (see Fig.~\ref{fig:1}). Their measured properties are listed in Table~\ref{tab:crsA}. Several of these sources have previously been identified as young stellar objects (YSOs) or candidate members of the embedded stellar population through infrared or X-ray observations (see Table~\ref{tab:crsA}), supporting the interpretation that a significant fraction of the detected radio sources trace magnetically active or wind-driving young stars associated with the Sh2-106 cluster.

We define the ``core'' as sources within $10''$ ($\sim0.08$\,pc at 1.65\,kpc) of S106\,IR. This radius corresponds approximately to the projected separation expected for stars ejected $\sim3500$\,yr ago with transverse velocities of up to $\sim20$\,km\,s$^{-1}$. Ten of the 23 detected compact sources lie within this core region, whereas the remaining 13 are distributed across the surrounding field. The concentration of compact radio sources toward the central $\sim0.1$\,pc suggests that many of these objects are physically associated with the embedded stellar cluster surrounding S106\,IR.

The compact sources span integrated flux densities between $\sim50\,\mu$Jy and $7.5$\,mJy. Their spectral indices range from $\alpha \sim -2.8$ to $\alpha \sim +1.7$ ($S_\nu \propto \nu^\alpha$). The core population exhibits a mixture of flat, positive, and negative spectral indices, consistent with a combination of partially optically thick free--free emission from ionized stellar winds or jets and nonthermal gyrosynchrotron emission associated with magnetically active young stars. The outer population is generally fainter and displays larger uncertainties in $\alpha$, consistent with a more distributed stellar population or unrelated background sources within the field of view.

The brightest source, S106\,IR (source \#5 in Table~\ref{tab:crsA}), shows a positive spectral index ($\alpha = 0.65 \pm 0.02$) and a partially resolved morphology, consistent with partially optically thick free--free emission from an ionized stellar wind. This result agrees with previous spectral index measurements of S106\,IR at similar frequencies; for example, \citet{masque2017} measured $\alpha = 0.75 \pm 0.08$, while \citet{bally2022} reported $\alpha \simeq 0.7$. Most compact sources are marginally resolved, with deconvolved sizes ranging from $\sim0.2''$ to $0.7''$, corresponding to physical scales of $\sim300$–$1200$\,AU.

In addition to the compact sources, we identify two slightly extended objects in the core region that exhibit bow-shock morphologies, with their apices pointing toward S106\,IR (see\,Fig.\,\ref{fig:1}c). This geometry suggests that they are externally photoionized by the massive star and may represent proplyd-like objects similar to those observed in other massive star-forming regions where external ionizing radiation interacts with circumstellar material \citep[e.g.,][]{henney1999,masque2014}. 

The compact radio sources detected in the central region therefore trace an embedded stellar population that is largely inaccessible to optical surveys because of the strong extinction toward Sh2-106. These objects provide a set of stellar tracers to probe the dynamical history of the central region.

\begin{table*}
\setlength{\tabcolsep}{1.0pt}
\renewcommand{\arraystretch}{1.2}
\centering
\scriptsize
 \caption{Compact radio sources in \ttar.}
    \begin{tabular}{ccccccccccccccccccccccccc}
         \hline
         \hline
\# & Other & Object& R.A. & Dec. & Size & {$S_{\rm int}$}& $S_{\rm peak}$ &
 {$S_{\rm int, LSB}$}& $S_{\rm peak,LSB}$ &  {$S_{\rm int, USB}$}& $S_{\rm peak,USB}$ & $\alpha$ \\
 & Name & Type &$20^{\rm h}27^{\rm m}[^{\rm s}]$&$37^{\circ}\,['\quad'']$& ($''\times''$; $^{\circ}$) &
($\mu$Jy) & ($\mu$Jy bm$^{-1}$)&($\mu$Jy) & ($\mu$Jy bm$^{-1}$)&($\mu$Jy) & ($\mu$Jy bm$^{-1}$)& 
 \\
 \hline
\multicolumn{10}{l}{Core region ($r\leq10''$)}\\ \cline{0-1}
1 & [OTN2006] 874 & YSO & 26.1930(3) & 22 43.395(2) & 0.16$\pm$0.02 $\times$ 0.08$\pm$0.02; 57$\pm$11 & 286 $\pm$ 14 & 214 $\pm$ 7 & 216 $\pm$ 15 & 235 $\pm$ 9 & 235 $\pm$ 17 & 210 $\pm$ 9 & 0.30 $\pm$ 0.35 \\
2 &  	[GFM2004b] S106 48 & X-ray & 26.4815(1) & 22 51.768(1) & 0.46$\pm$0.00 $\times$ 0.31$\pm$0.00; 52$\pm$1 & 4904 $\pm$ 33 & 1140 $\pm$ 6 & 6106 $\pm$ 46 & 1608 $\pm$ 10 & 3809 $\pm$ 41 & 1063 $\pm$ 9 & --1.67 $\pm$ 0.05 \\
3 & WISE J202726.50+372242.0  & YSO & 26.5709(5) & 22 42.225(5) & 0.42$\pm$0.02 $\times$ 0.33$\pm$0.01; 80$\pm$7 & 812 $\pm$ 32 & 195 $\pm$ 6 & 330 $\pm$ 27 & 185 $\pm$ 10 & 442 $\pm$ 37 & 147 $\pm$ 10 & 1.03\ $\pm$ 0.41 \\
4 & ... & ... & 26.5805(1) & 22 45.636(1) & \dots & 348 $\pm$ 12 & 323 $\pm$ 6 & 263 $\pm$ 14 & 305 $\pm$ 9 & 433 $\pm$ 19 & 344 $\pm$ 9 & 0.44 $\pm$ 0.01 \\
5 & S106\,IR & MYSO & 26.7715(1) & 22 47.687(1) & 0.240$\pm$0.003 $\times$ 0.141$\pm$0.002; 107$\pm$1 & 7467 $\pm$ 18 & 4072 $\pm$ 7 & 6996 $\pm$ 26 & 4124 $\pm$ 10 & 8395 $\pm$ 25 & 4676 $\pm$ 10 & 0.65 $\pm$ 0.02 \\
6 & MHO 4085 & Outflow & 26.8210(2) & 22 47.769(2) & 0.27$\pm$0.01 $\times$ 0.21$\pm$0.01; 51$\pm$8 & 764 $\pm$ 22 & 329 $\pm$ 7 & 885 $\pm$ 30 & 424 $\pm$ 10 & 925 $\pm$ 32 & 373 $\pm$ 10 & 0.16 $\pm$ 0.17 \\
7 &  	[OTN2006] 938 & YSO & 26.8274(6) & 22 45.287(3) & 0.45$\pm$0.02 $\times$ 0.22$\pm$0.01; 92$\pm$2 & 698 $\pm$ 27 & 208 $\pm$ 6 & 659 $\pm$ 30 & 309 $\pm$ 10 & 562 $\pm$ 34 & 206 $\pm$ 9 & --0.56 $\pm$ 0.27 \\
8 & [GFM2004b] S106 51 & X-ray & 26.8261(1) & 22 42.557(2) & 0.24$\pm$0.01 $\times$ 0.18$\pm$0.01; 38$\pm$7 & 871 $\pm$ 20 & 429 $\pm$ 7 & 471 $\pm$ 17 & 454 $\pm$ 10 & 751 $\pm$ 26 & 416 $\pm$ 10 & 1.65 $\pm$ 0.18 \\
9 & [GGC82] IRS 3S & IR & 26.8737(3) & 22 45.211(3) & 0.34$\pm$0.01 $\times$ 0.26$\pm$0.01; 76$\pm$6 & 823 $\pm$ 26 & 272 $\pm$ 7 & 823 $\pm$ 34 & 342 $\pm$ 10 & 509 $\pm$ 26 & 278 $\pm$ 10 & --1.70 $\pm$ 0.23 \\
10 & ... & ... & 27.0145(1) & 22 49.202(1) & 0.19$\pm$0.01 $\times$ 0.10$\pm$0.01; 11$\pm$3 & 1101 $\pm$ 16 & 730 $\pm$ 7 & 1088 $\pm$ 20 & 903 $\pm$ 10 & 1180 $\pm$ 24 & 721 $\pm$ 10 & 0.29 $\pm$ 0.10 \\ \cline{0-1}
 \multicolumn{10}{l}{Outer region ($r>10''$) }\\ \cline{0-1}
11 & 2MASS J20272329+3723261 & YSO & 23.3193(7) & 23 25.965(8) & \dots & 90 $\pm$ 13 & 77 $\pm$ 7 & 87 $\pm$ 18 & 80 $\pm$ 10 & 87 $\pm$ 15 & 88 $\pm$ 9 & 0.34 $\pm$ 0.56 \\
12 & 2MASS J20272380+3722452 & Star & 23.7904(6) & 22 44.970(4) & \dots & 122 $\pm$ 12 & 108 $\pm$ 6 & 101 $\pm$ 16 & 108 $\pm$ 9 & 100 $\pm$ 15 & 104 $\pm$ 9 & --0.11 $\pm$ 0.43 \\
13 & 2MASS J20272399+3723066 & Star & 23.9837(4) & 23 6.392(5) & \dots & 91 $\pm$ 10 & 98 $\pm$ 6 & 77 $\pm$ 15 & 87 $\pm$ 9 & 80 $\pm$ 12 & 108 $\pm$ 8 & 0.76 $\pm$ 046 \\
14 & ... & ... & 24.9150(18) & 22 6.884(14) & 0.31$\pm$0.07 $\times$ 0.17$\pm$0.07; 115$\pm$23 & 124 $\pm$ 21 & 54 $\pm$ 7 & 147 $\pm$ 35 & 57 $\pm$ 10 & 86 $\pm$ 26 & 46 $\pm$ 9 & --1.90 $\pm$ 1.36 \\
15 & ... & ... & 25.0314(12) & 22 8.106(12) & 0.23$\pm$0.06 $\times$ 0.20$\pm$0.07; 118$\pm$87 & 137 $\pm$ 20 & 66 $\pm$ 7 & 112 $\pm$ 25 & 69 $\pm$ 10 & 95 $\pm$ 21 & 65 $\pm$ 9 & --0.58 $\pm$ 1.11 \\
16 & ... & ... & 25.2366(9) & 22 05.630(12) & 0.39$\pm$0.04 $\times$ 0.26$\pm$0.03; 145$\pm$11 & 333 $\pm$ 29 & 95 $\pm$ 7 & 249 $\pm$ 30 & 124 $\pm$ 10 & 270 $\pm$ 41 & 75 $\pm$ 9 & 0.29 $\pm$ 0.69 \\
17 & [OTN2006] 816 & YSO & 25.3816(2) & 22 56.908(1) & 0.06$\pm$0.03 $\times$ 0.03$\pm$0.03; 129$\pm$78 & 283 $\pm$ 12 & 270 $\pm$ 6 & 238 $\pm$ 16 & 249 $\pm$ 9 & 319 $\pm$ 16 & 303 $\pm$ 9 & 1.04 $\pm$ 0.30 \\
18 & ... & ... & 25.6616(8) & 22 19.951(8) & 0.45$\pm$0.03 $\times$ 0.23$\pm$0.02; 129$\pm$4 & 449 $\pm$ 29 & 125 $\pm$ 6 & 346 $\pm$ 32 & 150 $\pm$ 10 & 246 $\pm$ 36 & 82 $\pm$ 9 & --1.21 $\pm$ 0.61 \\
19 & ... & ... & 25.9394(24) & 23 41.512(11) & 0.30$\pm$0.09 $\times$ 0.09$\pm$0.05; 83$\pm$15 & 82 $\pm$ 18 & 44 $\pm$ 7 & 77 $\pm$ 22 & 56 $\pm$ 10 & 35 $\pm$ 19 & 27 $\pm$ 8 & --2.79 $\pm$ 2.17 \\
20 & ... & ... & 25.9596(21) & 23 51.891(14) & 0.19$\pm$0.09 $\times$ 0.06$\pm$0.09; 107$\pm$47 & 53 $\pm$ 15 & 38 $\pm$ 7 & 70 $\pm$ 26 & 40 $\pm$ 10 & 27 $\pm$ 12 & 37 $\pm$ 8 & --3.37 $\pm$ 2.05 \\
21 & ... & ... & 26.0882(10) & 23 23.622(9) & 0.39$\pm$0.04 $\times$ 0.29$\pm$0.03; 79$\pm$15 & 371 $\pm$ 30 & 102 $\pm$ 6 & 294 $\pm$ 32 & 131 $\pm$ 10 & 267 $\pm$ 40 & 80 $\pm$ 9 & --0.34 $\pm$ 0.66 \\
22 & ... & ... & 27.0563(25) & 23 53.598(30) & 0.37$\pm$0.12 $\times$ 0.18$\pm$0.09; 41$\pm$30 & 95 $\pm$ 24 & 36 $\pm$ 7 & 71 $\pm$ 19 & 58 $\pm$ 9 & 86 $\pm$ 31 & 32 $\pm$ 8 & 0.68 $\pm$ 1.59 \\
23 & ... & ... & 27.6585(3) & 21 56.404(3) & \dots & 168 $\pm$ 12 & 153 $\pm$ 6 & 142 $\pm$ 15 & 163 $\pm$ 9 & 158 $\pm$ 17 & 149 $\pm$ 9 & -0.32 $\pm$ 0.29 \\
         \hline
    \end{tabular} \label{tab:crsA}\\
\end{table*}

\section{Discussions}

\subsection{The distance to Sh2-106}\label{sec:dist}

Distance estimates to Sh2-106 reported in the literature span a wide range. Early work commonly adopted a distance of $\sim600$\,pc based on UBVRI photometry of field stars and the identification of a marked increase in extinction along the line of sight \citep[e.g.,][]{staude1982}. Subsequent studies have noted, however, that this approach is sensitive to foreground structure and may primarily trace the distance to the first major dust layer in the direction of S106 rather than the distance to the nebula itself \citep[e.g.,][]{saito2009}.

More recent investigations argue that Sh2-106 is physically associated with the Cygnus~X molecular cloud complex, implying a significantly larger distance \citep[e.g.,][]{Schneider2007}. Trigonometric maser parallax measurements within the BeSSeL survey have provided direct geometric constraints on several star-forming regions in Cygnus~X. In particular, \citet{xu2013} measured a distance of $1.30 \pm 0.09$\,kpc toward Sh2-106 and proposed that the region lies in the forepart of the Cygnus~X complex. Subsequent parallax measurements of other Cygnus~X regions indicate that the complex is multi-layered, with major components located at distances of $\sim1.3$--1.5\,kpc \citep[e.g.,][]{rygl2012}.

Additional studies have estimated the distance using extinction–parallax techniques that combine Gaia data with stellar photometry. \citet{zucker2020} applied the so-called ``Wolf diagram'' or extinction breakpoint method to Gaia DR2 stars along the line of sight. In this approach, the cloud distance is inferred from the parallax at which the observed reddening or extinction increases abruptly due to dense dust associated with the target cloud. Using this method, Zucker et al. estimated a distance of $1.09 \pm 0.05$\,kpc to Sh2-106, a value subsequently adopted in several recent analyses of the region.

The extinction breakpoint technique has demonstrated utility for mapping distances to nearby molecular clouds in a homogeneous manner over large areas of the sky \citep[e.g.,][]{zucker2019,zucker2020} and generally agrees with independent distance indicators to within $\sim5$--10\% in regions where a single dominant cloud produces a clear extinction jump. However, the method has limitations in complex sightlines, particularly toward active star-forming regions in the Galactic plane, where multiple dust structures and spatial gradients can produce several extinction features along the line of sight. In such cases, identifying a unique breakpoint can be ambiguous and may preferentially trace nearer foreground dust layers rather than the true target cloud, especially if the extinction signature of the cloud itself is broad or spatially extended.

We also examined the recent 3D dust map of \citet{edenhofer2024}. The extinction structure toward Sh2-106 appears complex, with multiple increases along the line of sight rather than a single well-defined extinction jump. This suggests that extinction-break methods alone may not uniquely constrain the distance to the embedded stellar population in this direction.

Our Gaia DR3 astrometric analysis provides a new and independent geometric distance determination. By identifying a kinematically coherent population in the vicinity of S106\,IR and assessing the stability of the parallax signal as a function of selection radius, we find that the cluster parallax remains robust within $r \leq 3'$ and becomes increasingly affected by field contamination at larger radii. Adopting the $3'$ selection yields $\varpi_{\rm corr} = 0.607\pm0.013$\,mas, corresponding to $d = 1.65 \pm 0.04$\,kpc.

This cluster-based geometric distance is inconsistent with the $1.09 \pm 0.05$\,kpc extinction-break estimate of \citet{zucker2020}. While extinction-based methods effectively identify major foreground dust layers, they do not necessarily isolate the embedded stellar population in complex Galactic-plane sightlines. The Gaia result, instead, traces a kinematically coherent cluster that is spatially associated with S106\,IR. Consequently, interpretations of Sh2-106 that assumed a distance of 1.09\,kpc should be revisited.

The larger distance has direct implications for the proposed explosive expansion of the nebula. The dynamical age of $\sim3500$\,yr derived by \citet{bally2022} is based on the observed angular Hubble flow and is therefore largely independent of the adopted distance. However, the inferred physical parameters scale strongly with distance. For optically thin free--free emission, the ionized mass scales approximately as $D^{2.5}$, while velocities inferred from the observed angular expansion scale linearly with $D$, implying that the kinetic energy scales as $D^{4.5}$.

Adopting $d = 1.65$\,kpc instead of 1.09\,kpc increases the explosion energy by a factor of $(1.65/1.09)^{4.5} \approx 6.5$. Scaling the $\sim2\times10^{47}$\,erg ionized-gas kinetic energy estimate of \citet{bally2022} therefore yields $E_{\rm exp} \approx 1.3\times10^{48}$\,erg. Independent estimates of the kinetic energy of the Orion BN/KL explosion, dominated by its high-velocity molecular streamers, are of order $\sim10^{48}$\,erg \citep[e.g.,][]{bally2017,zapata2009}. Although the Sh2-106 estimate refers to the ionized component only, the revised value places Sh2-106 in the same order-of-magnitude energetic regime as BN/KL, albeit at a significantly older dynamical age ($\sim3500$\,yr versus $\sim550$\,yr). A quantitative comparison of the total (molecular + ionized) energy budgets will require dedicated measurements of the molecular component in Sh2-106.

Unlike Orion BN/KL, where high-velocity stellar ejecta provide direct evidence for a dynamical disintegration, no unambiguous runaway population is currently identified in Sh2-106. The revised Gaia cluster distance nevertheless enables a meaningful search for stellar-recoil signatures using the proper motions of cluster members. In the following subsection, we examine the Gaia DR3 astrometry of the identified cluster population to test whether any stars exhibit peculiar motions consistent with dynamical ejection from the vicinity of S106\,IR.

\subsection{Search for runaway stars in Sh2-106 using Gaia DR3 astrometry}

The availability of precise Gaia DR3 astrometry, together with the identification of a coherent stellar cluster in the vicinity of S106 IR, provides a natural opportunity to search for high-velocity stellar ejecta. This test is particularly motivated by the suggestion that Sh2-106 may host EMOs, analogous to the Orion BN/KL event, where a dynamical disintegration produced runaway stars with peculiar velocities of order $\sim10$ km s$^{-1}$ \citep{rodriguez2020}.

The Gaia DR3 proper motion of S106 IR ($\mu_{\alpha*} = -2.18 \pm 0.13$ mas yr$^{-1}$, $\mu_{\delta} = -5.86 \pm 0.15$ mas yr$^{-1}$) differs modestly from the mean cluster motion derived above. The resulting peculiar proper motion relative to the cluster centroid is $\Delta\mu_{\alpha*} = +0.17$ mas yr$^{-1}$ and $\Delta\mu_{\delta} = -0.66$ mas yr$^{-1}$, corresponding to a total peculiar motion of $\sim0.68$ mas yr$^{-1}$. At a distance of 1.65 kpc, this implies a transverse velocity of $\sim5$ km s$^{-1}$. The offset is detected at approximately the $3\sigma$ level when measurement uncertainties are accounted for.

The modest peculiar transverse velocity of S106\,IR ($\sim5$\,km\,s$^{-1}$) is well below the velocities observed for the runaway sources in Orion BN/KL, but it exceeds the typical internal velocity dispersion of young embedded clusters. In a dynamical interaction scenario, momentum conservation predicts that the most massive participant should acquire the smallest recoil velocity. Given a total system mass of $\sim23\,M_\odot$, the observed motion is therefore not inconsistent with participation in a past multi-body interaction, although it does not by itself constitute evidence for such an event.

While this velocity is well below those typically associated with classical runaway OB stars produced by dynamical few-body interactions ($\gtrsim20$ km s$^{-1}$; e.g., Poveda et al. 1967; Hoogerwerf et al. 2001), it exceeds the $\sim1$--3 km s$^{-1}$ internal velocity dispersion commonly observed in dynamically young clusters (e.g., Lada \& Lada 2003; Kuhn et al. 2019). The derived velocity is also comparable to the expected escape velocity of a moderately massive embedded cluster.

If S106\,IR participated in a past dynamical interaction or partial disintegration event, its relatively large mass (M$_{\rm Total}\sim23\,M_\odot$) would naturally limit the magnitude of its recoil velocity through momentum conservation, leading to a smaller acceleration compared to lower-mass participants. The observed velocity is therefore consistent with mild dynamical evolution but does not support a high-velocity runaway scenario.

\begin{figure}
    \centering
    \includegraphics[width=0.99\linewidth]{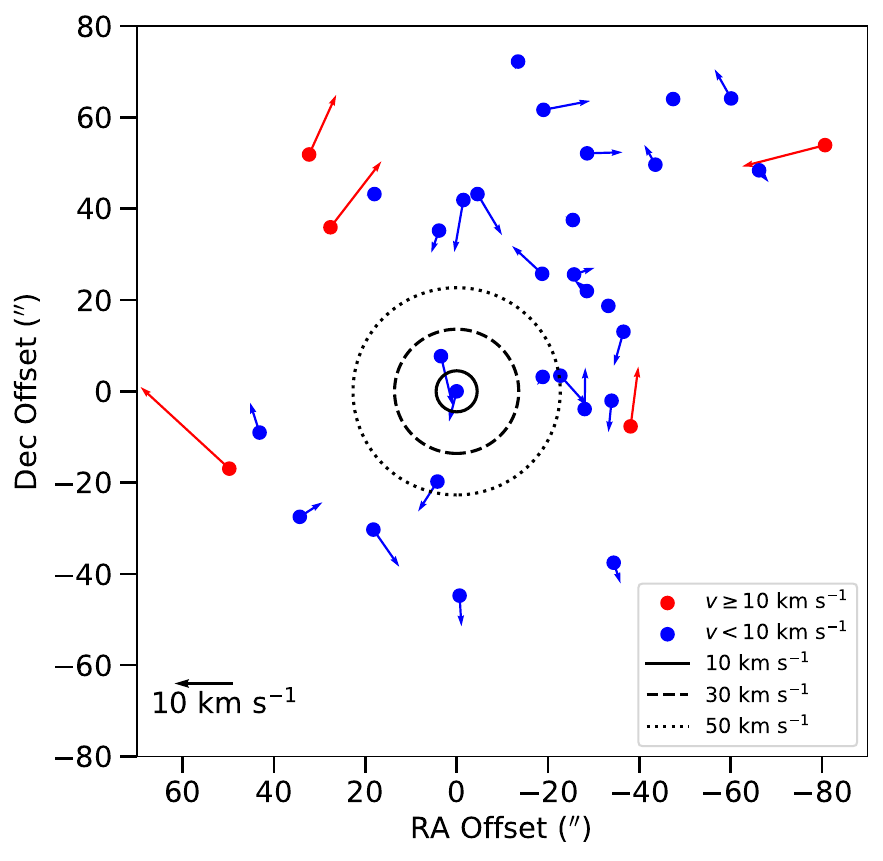}
    \caption{Spatial distribution and peculiar proper motions of Cluster 0 members relative to S106 IR. Red points mark stars with high peculiar transverse velocities ($>10$\,km\,s$^{-1}$), while blue points correspond to the remaining cluster members. Arrows indicate the peculiar proper motion vectors after subtraction of the cluster centroid motion. Concentric circles centered on S106 IR show the projected separations expected for stars ejected 3500 yr ago with purely tangential velocities of 10, 30, and 50 km s$^{-1}$ at a distance of 1.65 kpc. The scale arrow indicates 10 km s$^{-1}$.}
    \label{fig:runaways}
\end{figure}

We further searched for additional high-velocity cluster members using their peculiar proper motions relative to the cluster centroid (see Fig.~\ref{fig:runaways}). The resulting velocity vectors appear randomly distributed, and none of the high velocity ($>10$\,km s$^{-1}$) candidates exhibit motion vectors consistent with origin at S106\,IR, as would be expected for a recent dynamical disintegration event. 

Assuming a characteristic timescale of $\sim3500$ yr for a putative explosive event, stars ejected with transverse velocities of 10, 30, and 50 km s$^{-1}$ would today lie at projected separations of approximately 4.5$''$, 14$''$, and 23$''$, respectively, at a distance of 1.65 kpc. Even relatively high-velocity ejecta would therefore remain within $\sim30''$ of the cluster center. In the case of the Orion BN/KL disintegration, typical stellar velocities are of order $\sim10$\,km\,s$^{-1}$, implying expected separations of only a few arcseconds over similar timescales.

However, the innermost $\sim20''$ region surrounding S106 IR is heavily obscured, and Gaia detects only four sources within the area corresponding to the 50 km s$^{-1}$ scenario, including S106\,IR itself (see Fig.~\ref{fig:runaways}). The strong extinction in this region severely limits the completeness of the Gaia-based search for stellar ejecta. 

The absence of clear optical runaway stars, therefore, does not exclude the possibility of a past dynamical interaction, because the central $\sim20''$ region surrounding S106 IR is heavily obscured and incompletely sampled by Gaia. Deeply embedded ejecta would remain undetected at optical wavelengths. This limitation motivates the complementary search for embedded stellar tracers using radio observations, which are largely unaffected by extinction. In the following section, we examine the compact radio population of Sh2-106 and assess its relevance for testing the explosive-outflow hypothesis.

\subsection{Compact radio population}\label{sec:rpm}

The separation of the compact radio sources into a core ($r \leq 10''$) and an outer population ($r > 10''$) reveals distinct physical characteristics that are relevant to the explosive-outflow hypothesis. The core region contains ten compact sources within a projected radius of $\sim0.08$\,pc from S106\,IR. The surface density of compact radio sources within the core corresponds to $\sim500$ sources pc$^{-2}$, comparable to the radio source density observed in the Orion BN/KL region \citep[e.g.,][]{forbrich2016}.  The core population spans a wide range of spectral indices, including strongly negative ($\alpha < -1$),  flat, and positive values. Several sources exhibit partially resolved morphologies consistent with compact ionized winds, photoionized structures, or magnetically active young stars. The mixture of thermal and nonthermal spectral signatures indicates a dynamically young and embedded stellar population in the immediate vicinity of S106\,IR.

In contrast, the 13 compact sources located outside the  10$''$ radius are generally fainter and display larger uncertainties in their spectral indices. Their spatial distribution is more uniform across the field, and no concentration toward the proposed explosion center is evident. This outer population is consistent with a distributed cluster membership or unrelated background/foreground sources within the field of view. Their properties do not, by themselves, provide constraints on the dynamical history of the central region.

While the present radio observations do not directly test the explosive-outflow hypothesis, the identification of compact embedded radio sources provides a set of stellar tracers that can be followed with future multi-epoch radio interferometric observations. Because radio wavelengths are largely unaffected by the strong extinction toward the central region of Sh2-106, these sources offer the most promising avenue to measure proper motions of deeply embedded stars and to determine whether any high-velocity ejecta are present.

If Sh2-106 experienced a BN/KL-like dynamical disintegration, the most relevant stellar tracers would be those located within or emerging from the core region. Lower-mass participants in a few-body interaction would be expected to acquire larger recoil velocities than the massive primary and could remain embedded within the central $\sim0.1$\,pc over a $\sim3500$\,yr timescale. At a distance of 1.65\,kpc, a transverse velocity of 10\,km\,s$^{-1}$ corresponds to $\sim1.3$\,mas\,yr$^{-1}$, which is measurable over a multi-year baseline with current radio interferometric facilities.

Although the present observations do not reveal direct evidence for stellar ejecta, the existence of a compact embedded core population establishes a set of radio-bright stellar tracers that can be monitored with future multi-epoch radio astrometry. Because radio wavelengths penetrate the strong extinction toward the central region of Sh2-106, such measurements provide the most promising route to detect or rule out high-velocity stellar ejecta associated with a past dynamical interaction. Combined with the revised Gaia cluster distance presented in this work, these observations would enable a definitive test of whether Sh2-106 represents an older analogue of the Orion BN/KL explosive event or instead reflects a distinct physical mechanism.

\section{Conclusions} \label{sec:conclusions}

We have presented high-angular-resolution VLA observations at 5\,GHz together with a Gaia DR3 astrometric analysis of the Sh2-106 region. Our main findings are:

\begin{enumerate}

\item[-] Using Gaia DR3 data, we identify a kinematically coherent stellar- population associated with S106 IR and derive a cluster parallax of $\varpi_{\rm corr} = 0.607\pm0.013$\,mas, corresponding to a distance of $1.65 \pm 0.04$\,kpc. This geometric distance is inconsistent with the commonly adopted 1.09\,kpc extinction-break estimate and places Sh2-106 firmly within the Cygnus~X complex.

\item[-] At this revised distance, the kinetic energy of the expanding ionized nebula increases by a factor of $\sim6.5$ relative to previous estimates, yielding $E_{\rm exp} \approx 1.3\times10^{48}$\,erg when scaling the values reported by \citet{bally2022}. Although this estimate refers to the ionized component only, Sh2-106 now occupies the $\sim10^{48}$\,erg energetic regime characteristic of the Orion BN/KL explosive event, albeit at a significantly older dynamical age ($\sim3500$\,yr versus $\sim550$\,yr).

\item[-] No unambiguous high-velocity stellar ejecta are identified in Gaia DR3. The massive source S106\,IR exhibits a modest peculiar transverse velocity of $\sim5$\,km\,s$^{-1}$ relative to the cluster centroid, well below classical runaway values. However, strong extinction in the innermost region limits the completeness of the optical search.

\item[-] Our VLA data reveal ten compact radio sources within the central region, including S106 IR and several embedded objects. These sources provide the first embedded stellar tracers suitable for future multi-epoch radio astrometry aimed at testing the dynamical decay hypothesis.

\end{enumerate}

The revised Gaia distance elevates Sh2-106 to the same order-of-magnitude energetic regime as BN/KL. While no clear stellar recoil signatures are currently detected, the presence of an embedded radio population opens a direct observational path to test whether Sh2-106 represents an older analogue of the Orion BN/KL dynamical event or a physically distinct explosive phenomenon.

\begin{acknowledgements}
S.A.D. acknowledges the M2FINDERS project from the European Research
Council (ERC) under the European Union's Horizon 2020 research and innovation programme
(grant No 101018682).
The National Radio Astronomy Observatory is a facility of the National Science Foundation operated under cooperative agreement by Associated Universities, Inc.
\end{acknowledgements}

\bibliographystyle{aa_url} 
\bibliography{references.bib}

\appendix

\section{Robustness of the Gaia cluster identification}\label{app:g}

\begin{figure}[h]
    \centering
    \includegraphics[width=1.0\linewidth]{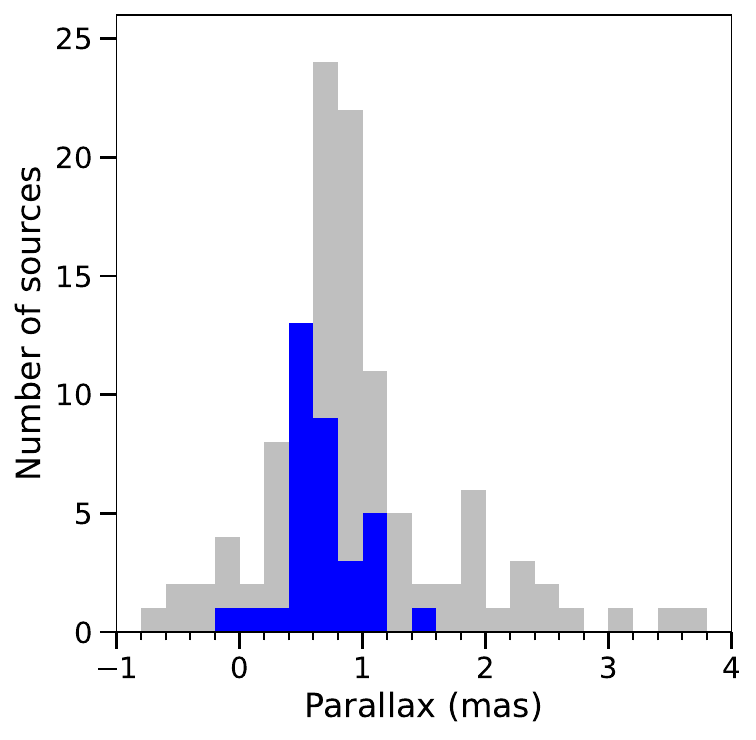}
    \caption{Parallax distribution of Gaia sources within $5'$ of Sh2-106. Stars identified as Cluster~0 members by the DBSCAN analysis (orange) are compared with field stars (blue). The cluster members show a clear peak near $\varpi \approx 0.6$\,mas, corresponding to a distance of $d \approx 1.65$\,kpc.}
    \label{fig:g3}
\end{figure}

Figure~\ref{fig:g3} shows the parallax distribution of the stars identified as members of Cluster~0. The distribution peaks near $\varpi \sim 0.6$\,mas, consistent with the value adopted in the main analysis.

\begin{figure}
    \centering
    \includegraphics[width=1.0\linewidth]{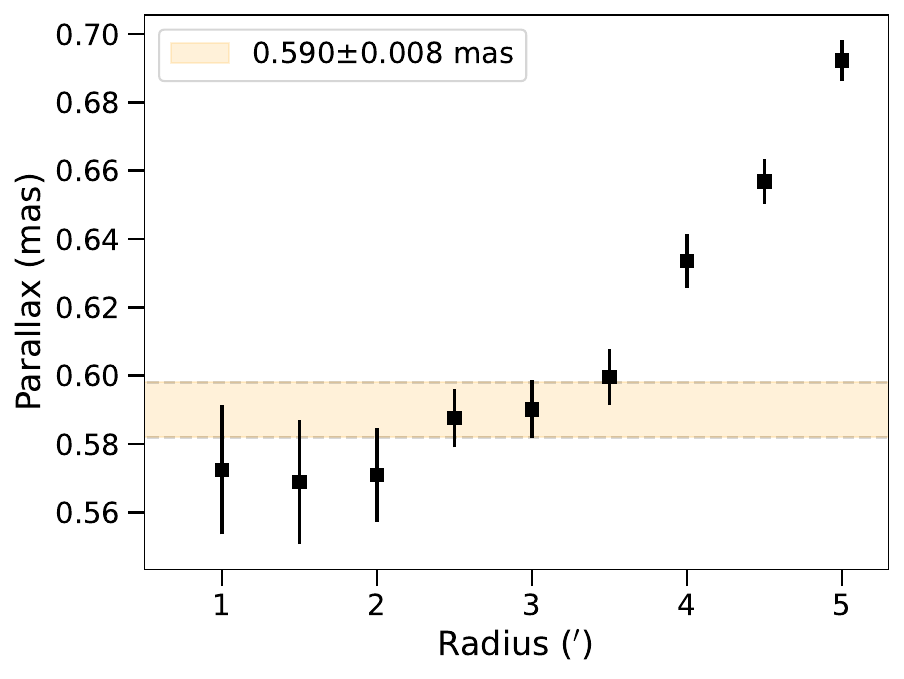}
    \caption{Mean parallax of Cluster~0 as a function of the selection radius used in the DBSCAN clustering analysis. Error bars indicate the uncertainty of the weighted mean. The mean parallax increases beyond $3'$, indicating growing contamination from foreground stars at larger radii. The $3'$ sample is therefore adopted for the cluster distance estimate.}
    \label{fig:g4}
\end{figure}

To evaluate the effect of field contamination, the clustering analysis was repeated for different selection radii. The resulting mean parallax values are shown in Fig.~\ref{fig:g4}. While the proper-motion centroid remains stable across all radii, the mean parallax increases beyond $3'$ as contamination from foreground stars becomes significant. This supports the adoption of the $3'$ sample for the cluster distance estimate.


\end{document}